\documentclass[12pt]{article}
\usepackage{a4}
\usepackage{amsmath}
\usepackage{amsfonts}
\usepackage{amsmath}
\usepackage{latexsym}
\usepackage{amsfonts}
\begin{document}
\oddsidemargin 6pt\evensidemargin 6pt\marginparwidth
48pt\marginparsep 10pt

\renewcommand{\thefootnote}{\fnsymbol{footnote}}
\thispagestyle{empty}

\noindent    \hfill  October    2006 \\

\noindent \vskip3.3cm
\begin{center}

{\Large\bf Conformal coupling of the scalar field with gravity in
higher dimensions and invariant powers of the Laplacian}
\bigskip\bigskip\bigskip

{\large Ruben Manvelyan ${}^{\dagger \ddag}$ and D. H.
Tchrakian}$^{\dagger \star}$\\
\medskip
$^{\dagger}${\small\it Department of Mathematical Physics, National
University of Ireland Maynooth,} {\small\it Maynooth, Ireland}
\\\medskip
$^{\star}${\small\it School of Theoretical Physics -- DIAS, 10
Burlington Road, Dublin 4, Ireland }\\
\medskip
${}^{\ddag}${\small\it Yerevan Physics Institute\\ Alikhanian Br.
Str.
2, 0036 Yerevan, Armenia}\\
\medskip
{\small\tt e-mails manvel@physik.uni-kl.de, tigran@thphys.nuim.ie}
\end{center}

\bigskip 
\begin{center}
{\sc Abstract}
\end{center}

The hierarchy of conformally coupled scalars  with the increasing
scaling dimensions $\Delta_{k}=k-d/2$, $k=1,2,3,\dots $ connected
with the $k$-th Euler density  in the corresponding space-time
dimensions $d\geq 2k$ is proposed. The corresponding conformal
invariant Lagrangian with the $k$-th power of Laplacian for the
already known cases $k=1,2$ is reviewed, and the subsequent case of
$k=3$ is completely constructed and analyzed.

\newpage

\section{Introduction}

\quad Our aim in this work is the construction of a hierarchy of
conformally invariant Lagrangians in spacetime dimensions $d\geq
2k$, describing the nonminimal coupling of gravity with a scalar
field whose conformal dimension is $\Delta_{(k)}=k-d/2$. A
remarkable feature of these systems is the appearance of the $k$-th
Euler density $E_{(k)}$ in the $k$-th member of this hierarchy. The
$k=1,2$ cases are known, and here we supply the $k=3$ case
concretely, suggesting the arbitrary $k$ case.

 The conformal
coupling of a scalar field with gravity in different dimensions has
been a subject of interest in quantum field theory in curved
spacetimes~\cite{BD}. In recent years it has attracted special
attention in the context of new developments in the area of
$AdS/CFT$ \cite{mald} correspondence, and in investigations of
higher order and higher spin gravitating systems in
general~\cite{OP}. Conformally invariant field theories in higher
dimensions are particularly interesting because they present a
universal tool for investigations of their quantum properties, such
as conformal or trace anomalies~\cite{anom}. Another important
properety of conformally invariant theories in arbitrary dimensions
is, that the method of dimensional regularisation can be employed as
a conformally invariant regularisation in higher dimensions for the
construction of anomalous effective actions \cite{Schwim}. Note also
that in connection with higher spin gauge field interactions with a
scalar field, this coupling and Weyl invariance itself, can be
generalized~\cite{RM}.

In this article we propose a hierarchy of such couplings of gravity
to scalar fields with increasing scaling dimensions parameterized by
a natural number $k$, and living in all space-time dimensions $d\geq
2k$. Actually this hierarchy corresponds to the conformally
invariant $k$-th power of the Laplacian acting on a scalar field
with conformal dimension $\Delta_{(k)}=k-d/2$, in spacetime
dimensions $d\geq 2k$. From the other hand we propose the connection
between this hierarchy and the $k$-th Euler density $E_{(k)}$ lifted
to spacetime dimensions greater than $2k$. For completeness, we
verify this proposal in the well known text book case of
$k=1$~\cite{BD}. We then turn to the known case in
$d=4$~\cite{Riegert,FT}, and the fourth order conformally covariant
operator in dimension $d\geq 4$ obtained in \cite{pan,es} long ago,
which provides us with a further check of our proposal, now
involving the second Euler density $E_{(2)}$. Finally in the last
section we perform the new calculation of the locally Weyl invariant
third power of the Laplacian in spacetime dimensions $d\geq 6$, or
in another words we construct a conformally invariant action for the
scalar with conformal dimension $3-d/2$ coupled with gravity. In all
three cases we have found the corresponding Euler density $E_{(k)}$
as part of the invariant action, proportional to the first order of
$\Delta_{(k)}$, and without derivatives. Taking into account the
rather technical character of this article we devote a substantial
section, Section {\bf 2}, with a more or less complete technical
setup and all the formulas which we have used in our calculations.

\section{Notations and Conventions}
We work in a $d$ dimensional curved space and use the following
conventions for covariant derivatives and curvatures:
\begin{eqnarray}
  \nabla_{\mu}V^{\rho}_{\lambda}&=& \partial_{\mu}V^{\rho}_{\lambda}+
  \Gamma^{\rho}_{\mu\sigma}V^{\sigma}_{\lambda}-\Gamma^{\sigma}_{\mu\lambda}V^{\rho}_{\sigma} , \\
  \Gamma^{\rho}_{\mu\nu} &=& \frac{1}{2} g^{\rho\lambda}\left(\partial_{\mu}g_{\nu\lambda}+
  \partial_{\nu}g_{\mu\lambda} - \partial_{\lambda}g_{\mu\nu}\right) , \\
  \left[\nabla_{\mu} , \nabla_{\nu}\right]V^{\rho}_{\lambda} &=&
  R^{\quad\,\,\rho}_{\mu\nu \sigma}V^{\sigma}_{\lambda}
  -R^{\quad\,\,\sigma}_{\mu\nu \lambda}V^{\rho}_{\sigma} ,\\
  R^{\quad\,\,\rho}_{\mu\nu \lambda}&=& \partial_{\mu}\Gamma^{\rho}_{\nu\lambda}
  -\partial_{\nu}\Gamma^{\rho}_{\mu\lambda}+\Gamma^{\rho}_{\mu\sigma}\Gamma^{\sigma}_{\nu\lambda}
  -\Gamma^{\rho}_{\nu\sigma}\Gamma^{\sigma}_{\mu\lambda} ,\\
  R_{\mu\lambda}&=& R^{\quad\,\,\rho}_{\mu\rho\lambda}\quad , \quad
  R=R^{\,\,\mu}_{\mu} .
\end{eqnarray}
The corresponding local conformal transformations (Weyl rescalings)
\begin{eqnarray}
  \delta g_{\mu\nu}&=&2\sigma(x) g_{\mu\nu} , \quad\
  \delta g^{\mu\nu} = -2\sigma(x) g^{\mu\nu} ,\\
   \delta\Gamma^{\lambda}_{\mu\nu}&=& \partial_{\mu}\sigma\delta^{\lambda}_{\nu}
  +\partial_{\nu}\sigma\delta^{\lambda}_{\mu}-
  g_{\mu\nu}\partial^{\lambda}\sigma , \label{christ}\\
   \delta R^{\quad\,\,\rho}_{\mu\nu \lambda}&=&\nabla_{\mu}
  \partial_{\lambda}\sigma\delta^{\rho}_{\nu}-
  \nabla_{\nu}\partial_{\lambda}\sigma\delta^{\rho}_{\mu}+
  g_{\mu\lambda}\nabla_{\nu}\partial^{\rho}\sigma
  - g_{\nu\lambda}\nabla_{\mu}\partial^{\rho}\sigma ,\\
  \delta R_{\mu\lambda}&=&(d-2)\nabla_{\mu}\partial_{\lambda}\sigma +
g_{\mu\lambda}\Box
\sigma ,\\
 \delta R&=& -2\sigma R + 2(d-1)\Box \sigma\,,
\end{eqnarray}
are first oder in the infinitesimal local scaling parameter
$\sigma$.

We then introduce the Weyl ($W$) and Schouten ($K$) tensors, as well as
the scalar $J$
\begin{eqnarray}
  R_{\mu \nu } &=& (d - 2)K_{\mu \nu }  + g_{\mu \nu}J , \quad J=\frac{1}{2(d - 1)}R\,\,, \\
  W_{\mu \nu \lambda }^{\quad\,\,\rho}   &=& R_{\mu \nu \lambda } ^\rho   - K_{\mu \lambda }
 \delta _\nu  ^\rho   +
 K_{\nu \lambda } \delta _\mu  ^\rho   - K_\nu  ^\rho  g_{\mu \lambda }  +
 K_\mu  ^\rho  g_{\nu \lambda } \,\,, \label{12}\\
  \delta K_{\mu \nu } &=& \nabla _\mu  \partial _\nu
 \sigma \,,\,\,\,\, \label{13}\\ \delta J&=&-2\sigma J + \Box
 \sigma\,,\,\,\,\,\,\label{deltaJ}\\
 \delta W_{\mu \nu \lambda }^{\quad\,\,\rho}&=& 0\,,
\end{eqnarray}
which are more convenient because their conformal transformations
are "diagonal".

To describe the Bianchi identity with these tensors, we have to
introduce the so called Cotton tensor
\begin{eqnarray}
  C_{\mu \nu \lambda } &=& \nabla _\mu  K_{\nu \lambda }  - \nabla _\nu
 K_{\mu \lambda } \,\,,
  \label{bi1}\\ \delta C_{\mu \nu \lambda }
 &=&  - \partial _\alpha  \sigma W_{\mu \nu \lambda
 }^{\quad\,\,\alpha}\,\,\,\,,\,\,\,\,\,  C_{[\mu \nu \lambda ] = 0\,\,.}
\end{eqnarray}
All important properties of these tensors following from the Bianchi
identity can then be listed as
\begin{eqnarray}
  \nabla _{[\alpha } W_{\mu \nu ]\lambda }^{\quad\,\,\,\rho} &=& g_{\lambda [\alpha }
  C_{\mu \nu ]}^{\quad\,\,\rho}
   - \delta _{[\alpha }^\rho  C_{\mu \nu ]\lambda } \,\,,\, \label{18}\\
 \nabla _\alpha  W_{\mu \nu \lambda }^{\quad\,\,\alpha}&=& \left( {3
- d}\right)C_{\mu \nu \lambda } \,\,\,,\\
 \nabla^{\mu}K_{\mu\nu}&=&\partial_{\nu}J\,,\label{bi2}\\
 C_{\mu \nu }^{\quad\,\nu} &=& 0\,\,,\quad\quad
\nabla^{\lambda}C_{\mu\nu\lambda}=0\,.
\end{eqnarray}
Finally we introduce the last important conformal
object in the above listed hierarchy, namely the symmetric and traceless
Bach tensor
\begin{eqnarray}
B_{\mu\nu}&=&\nabla^{\lambda}C_{\lambda\mu\nu}
  +K^{\lambda}_{\alpha}W_{\lambda\mu \nu  }^{\quad\,\,\alpha} ,
\end{eqnarray}
whose conformal transformation and divergence are expressed in terms
of the Cotton and the  Schouten tensors as follows
\begin{eqnarray}
    \delta B_{\mu\nu}&=&-2\sigma B_{\mu\nu}+(d-4)\nabla^{\lambda}\sigma
  \left( C_{\lambda\mu\nu}+C_{\lambda\nu\mu}\right) ,\label{bachtr}\\
   \nabla^{\mu}B_{\mu\nu}&=&(d-4)C_{\alpha\nu\beta}K^{\alpha\beta} .
  \end{eqnarray}
Note that only in four dimensions is the Bach tensor
conformally invariant and divergenceless.

This basis of $B,C,K,J,W$ tensors is all we need to construct any
conformally invariant object in arbitrary dimensions. Finally for any
scalar $f^{\Delta}(x)$ with arbitrary scaling dimension $\Delta$ we
can easily derive the following important relations
\begin{eqnarray}
   \delta\left(\nabla_{\mu}\partial_{\nu}f^{\Delta}\right)
  &=&\Delta\sigma \nabla_{\mu}\partial_{\nu}f^{\Delta}
  +\Delta f^{\Delta} \nabla_{\mu}\partial_{\nu}\sigma
  +(\Delta-1)\partial_{(\mu}\sigma\partial_{\nu)}
  f^{\Delta}+g_{\mu\nu}\partial^{\lambda}\sigma\partial_{\lambda}f^{\Delta} ,\quad\quad  \\
   \delta\left(\Box f^{\Delta}\right)&=&(\Delta-2)\sigma\Box
  f^{\Delta}+\Delta f^{\Delta}\Box \sigma +(d+2\Delta-2)
  \partial^{\lambda}\sigma\partial_{\lambda}f^{\Delta} ,\label{lap}
\end{eqnarray}
by using the
transformation (\ref{christ}) for Christoffel symbols.

\section{Hierarchies of conformal scalars and Euler densities}

\quad In this section we introduce the hierarchy of scalar fields
$\varphi_{(k)}$, where $k=1,2,3,\dots$ with the corresponding
scaling dimensions and infinitesimal conformal transformations
\begin{eqnarray}\label{trans}
   \Delta_{(k)}&=&k-d/2\,,\\
    \delta \varphi_{(k)}:&=&\Delta_{(k)}\sigma\varphi_{(k)} .
\end{eqnarray}
Each of these exist in the spacetime dimensions $d\geq 2k$, and with
the minimal dimension vanishing, $\Delta_{(k)}=0$ when $d=2k$.

Let us now introduce the hierarchy of the Euler densities~\footnote{Note
that the usual Einstein--Hilbert Lagrangian in $d$ dimensions is
the $k=1$ member of this hierarchy of gravitational Lagrangians.}
\begin{eqnarray}
  && E_{(k)}:=\frac{1}{2k (d-2k)!}\delta^{\alpha_{1}
  \dots\alpha_{d-2k}\mu_{1}\mu_{2}\dots\mu_{2k-1}\mu_{2k}}_{\alpha_{1}
  \dots\alpha_{d-2k}\nu_{1}\nu_{2}\dots\nu_{2k-1}\nu_{2k}}
  R^{\nu_{1}\nu_{2}}_{\mu_{1}\mu_{2}}\dots R^{\nu_{2k-1}\nu_{2k}}_{\mu_{2k-1}
  \mu_{2k}} .\label{Ek}
\end{eqnarray}
This set of objects exist as Lagrangians in space time dimensions
$d\geq 2k$, but for the minimal case $d=2k$, $E_{k}$ is a total
divergenece such that its integral is a
topological invariant, the Euler characteristic. In these dimensions
$E_{k}$ trivialize as Lagrangians but describe the topological
part of the trace anomaly in the corresponding even space-time
dimension $2k$.

The idea of this article is the following observation: \emph{There should
be a one to one correspondence between the conformally coupled scalars
$\varphi_{(k)}$ and the Euler densities $E_{(k)}$}.

Our first step in proving this is to start from the action of the well
known non minimal conformally coupled scalar in the space-time dimension
$d$ and with conformal dimension $\Delta_{1}=1-d/2$
\begin{eqnarray}
  && S_{(1)}=\frac{1}{2}\int d^{d}x\sqrt{g}\left\{
  g^{\mu\nu}\partial_{\mu}\varphi_{(1)}\partial_{\nu}
  \varphi_{(1)}-\frac{d-2}{4(d-1)}R\varphi^{2}_{(1)}\right\} .\label{act1}
\end{eqnarray}
We first see that the second term without derivatives and proportional
to the scaling dimension can be written in the form
$-\frac{d-2}{4(d-1)}R=\Delta_{(1)}J$. After that the proof of the
conformal invariance of the action (\ref{act1}) becomes trivial: We
write (\ref{lap}) for $\Delta=\Delta_{(1)}$ and use
(\ref{deltaJ}), from which it follows that
$\delta\left[\sqrt{g}\varphi_{(1)}\left(\Box-\Delta_{(1)}J\right)\varphi_{(1)}\right]=0$. We next evaluate (\ref{Ek}) for $k=1$
\begin{equation}\label{e1}
E_{(1)}=2(d-1)J \, .
\end{equation}
Finally we see that (\ref{act1}) can be rewritten in the form
\begin{eqnarray}
  && S_{(1)}=\frac{1}{2}\int d^{d}x\sqrt{g}
  \left\{-\varphi_{(1)}\Box\varphi_{(1)}
  +\frac{\Delta_{(1)}}{2(d-1)}E_{(1)}\varphi^{2}_{(1)}\right\}\,. \label{act11}
\end{eqnarray}
We now see that \emph{derivative independent part of the conformally
invariant action is proportional to the scaling dimension times the first
Euler density}. Note again that both objects degenerate in minimal
dimension $d=2$ where the Laplacian itself is conformally invariant
and the Euler density describes the topological invariant, which is the
two dimensional trace anomaly.

The next step in our considerations is the $k=2$ case. Again this
higher derivative action (or 4-th order conformal invariant
operator) is known since many years~\cite{Riegert,FT} for dimension
$4$ as well as for general $d$~\cite{pan,es}. All this is presented
in \cite{Erd} where many of the invariant objects are considered. In
our work, we rederived this Lagrangian just applying the Noether
procedure to the local conformal variation of the following suitable
object
\begin{equation}\label{k21}
    S_{(2)}^{0}=\frac{1}{2}\int d^{d}x\sqrt{g}\left(\widehat{D}_{(2)}\varphi_{(2)}\right)^{2}
    ,
\end{equation}
whose Weyl transformation includes only the first derivatives of
the parameter. In (\ref{k21}) and
thereafter, we use the notation
\begin{eqnarray}\label{dk}
    \widehat{D}_{(k)}&:=&\Box - \Delta_{(k)}J , \quad k=1,2,3,\dots \,\,,\\
    \delta\left(\widehat{D}_{(k)}\varphi_{(k)}\right)&=&(\Delta_{(k)}-2)\widehat{D}_{(k)}
    \varphi_{(k)} +
    2(k-1)\partial^{\mu}\sigma\partial_{\mu}\varphi_{(k)} ,\\
    \widehat{D}^{(k)}_{\mu\nu}&:=&\nabla_{\mu}\partial_{\nu}-
    \Delta_{(k)}K_{\mu\nu}\,,\quad
    g^{\mu\nu}\widehat{D}^{(k)}_{\mu\nu}=\widehat{D}_{(k)}\,,\\
    \delta\left(\widehat{D}^{(k)}_{\mu\nu}\varphi_{(k)}\right)&=&
    \Delta_{(k)}\sigma\widehat{D}^{(k)}_{\mu\nu}\varphi_{(k)}+(\Delta_{(k)}-1)\partial_{(\mu}\sigma\partial_{\nu)}
  \varphi_{(k)}+g_{\mu\nu}\partial^{\lambda}\sigma\partial_{\lambda}\varphi_{(k)} .\quad\quad
\end{eqnarray}
Performing the functional integration of the Weyl variation of the
(\ref{k21}) is now just a matter of some partial integration,
elimination of the second derivatives of $\sigma$ using
(\ref{13}),(\ref{deltaJ}) and cancelation of terms linear in
$\partial\sigma$ using the Bianchi identity (\ref{bi2}). It should
be noted here that all these types of calculations could instead be
performed using the powerful method proposed in \cite{oraf}.  Here
we presented only the direct Noether procedure because that will be
more suitable for us in the next section. After all these
manipulations we arrive at the following action
\begin{eqnarray}
  S_{(2)}^{1}&=&\frac{1}{2}\int d^{d}x\sqrt{g}\Big\{\left(\widehat{D}_{(2)}
  \varphi_{(2)}\right)^{2} +4K^{\mu\nu}\partial_{\mu} \varphi_{(2)}
  \partial_{\nu}\varphi_{(2)} - 2 J\partial^{\mu}
  \varphi_{(2)}\partial_{\mu} \varphi_{(2)}\nonumber\\
  & & +2\Delta_{(2)}\left(K^{2}-J^{2}\right)\varphi^{2}_{(2)}\Big\}\label{s2}
\end{eqnarray}
Then after some work we can evaluate $E_{(2)}$ using (\ref{Ek}) and
(\ref{12}) to be
\begin{eqnarray}
  E_{(2)}&=& W^{2}-4(d-3)(d-2)\left(K^{2}-J^{2}\right)\,. \label{E2}
\end{eqnarray}
W see that the $\varphi^{2}_{(2)}$ term in
(\ref{s2}) which is linear in $\Delta_{(2)}$, is proportional to the
Weyl tensor independent part of the Euler
density. The other term without derivatives is proportional to
$\Delta_{(2)}^{2}$. This noninvariant part of the four dimensional
trace anomaly arises in $AdS/CFT$ \cite{HS} and carries the name
"holographic", and corresponds to the maximally supersymmetric gauge
theory on the boundary of $AdS_{4}$.

The combination
\begin{equation}\label{w2}
    -\frac{1}{2}\int
    d^{d}x\sqrt{g}\left\{\frac{\Delta_{2}}{2(d-3)(d-2)}W^{2}\varphi^{2}_{(2)}\right\}
    ,
\end{equation}
on the other hand is also conformally invariant and can be added to (\ref{s2}) at no cost. This leads us to our final result
\begin{eqnarray}
  S_{(2)}^{E}&=&\frac{1}{2}\int d^{d}x\sqrt{g}
  \left\{\varphi_{(2)}\Box^{2}\varphi_{(2)}
 -2\Delta_{(2)}J\varphi_{(2)}\Box\varphi_{(2)}
  + \Delta_{(2)}^{2}J^{2}\varphi^{2}_{(2)} \nonumber\right.\\
 &-& \left. 2 J\partial^{\mu}
  \varphi_{(2)}\partial_{\mu} \varphi_{(2)}
  +4K^{\mu\nu}\partial_{\mu} \varphi_{(2)} \partial_{\nu}\varphi_{(2)}
   -\frac{\Delta_{(2)}}{2(d-3)(d-2)}E_{(2)}\varphi^{2}_{(2)}
   \right\},\,\label{se}
\end{eqnarray}
confirming our main observation in the $k=2$ case.
\section{The $\Delta_{3}=3-d/2$ case}
To confirm our main observation, verified for $k=1,2$ above, and present
it as an assertion for general $k$, we need to carry out this verification
in the next nontrivial case of $k=3$. This is the content of the present
Section, which consists of the explicit calculation of the conformally
invariant action analogous to (\ref{act11}) and (\ref{se}) for $k=1,2$.
In this case we will follow again the same strategy.

Taking into account that $\widehat{D}_{(3)}\varphi_{(3)}$ scales as an
object with the dimension $\Delta_{(1)}=\Delta_{(3)}-2$ we start
from the following initial Lagrangian
\begin{equation}\label{ks0}
    S_{(3)}^{0}=-\frac{1}{2}\int
    d^{d}x\sqrt{g}\left\{\widehat{D}_{(3)}\varphi_{(3)}
    \left(\widehat{D}_{(3)}+2J\right)\widehat{D}_{(3)}\varphi_{(3)}\right\}
    ,
\end{equation}
with the more or less simple Weyl variation
\begin{eqnarray}
  &&\delta S_{(3)}^{0}=-\int d^{d}x\sqrt{g}
  \left\{4\widehat{D}_{(3)}\varphi_{(3)}\left(\Delta_{(3)}
  \varphi_{(3)}\partial^{\lambda}\sigma\partial_{\lambda}J +4(\Delta_{(3)}-2)
  K^{\mu\nu}\partial_{\mu}\sigma\partial_{\nu}\varphi_{(3)}\right)\right.\quad\nonumber\\
  &&\left. -2\widehat{D}_{(3)}\varphi_{(3)}\left(\widehat{D}_{(3)}\varphi_{(3)}
  \delta J -4\widehat{D}^{(3)}_{\mu\nu}
  \varphi_{(3)}\delta K^{\mu\nu} -
  2\partial_{\lambda}\varphi_{(3)}\partial^{\lambda}\delta
  J -2
  \Delta_{(3)}\delta(K^{2})\varphi_{(3)}\right)\right\}.\quad\quad\label{1st}
 \end{eqnarray}
The second line in (\ref{1st}) can be integrated adding to the
$S_{(3)}^{0}$ the following terms
\begin{eqnarray}
  S_{(3)}^{1}&=&-\int d^{d}x\sqrt{g}\left\{2(\widehat{D}_{(3)}\varphi_{(3)})^{2}
  J
  -8\widehat{D}_{(3)}\varphi_{(3)}\widehat{D}^{(3)}_{\mu\nu}
  \varphi_{(3)}K^{\mu\nu}\right.\nonumber\\
  & &\left.-
  4\widehat{D}_{(3)}\varphi_{(3)}\partial_{\lambda}\varphi_{(3)}\partial^{\lambda}
  J -4\Delta_{(3)}
  \widehat{D}_{(3)}\varphi_{(3)}K^{2}\varphi_{(3)}\right\} .\label{44}
\end{eqnarray}
Writing the variation of the $ S_{(3)}^{0+1}$ is rather more
complicated. First we should separate the Laplacians from
$\Delta_{(3)}J$ in the terms with $\widehat{D}_{(3)}\varphi_{(3)}$,
then, performing some partial integrations we redistribute
derivatives and separate the terms
$\partial_{\mu}\varphi_{(3)}\partial_{\nu}\varphi_{(3)}$,\,
$\partial_{\mu}\varphi_{(3)}\partial^{\mu}\varphi_{(3)}$ and
$\varphi_{(3)}^{2}$, that are irreducible under partial integration
. After some manipulations, using (\ref{bi1}) and Bianchi
identities, we obtain
 \begin{eqnarray}
   && \delta S_{(3)}^{0}+\delta S_{(3)}^{1}=-\delta S_{(3)}^{2}-\delta S_{(3)}^{\Delta_{(3)}}\nonumber\\
   && +
   \int d^{d}x\sqrt{g}\left\{16C^{\lambda\mu\nu}\partial_{\lambda}
   \sigma\partial_{\mu}\varphi_{(3)}\partial_{\nu}
   \varphi_{(3)}+24\Delta_{(3)}C^{\lambda\mu\nu}\partial_{\lambda}
   \sigma K_{\mu\nu}\varphi_{(3)}^{2}\right\} ,\label{cotton}
 \end{eqnarray}
where
\begin{eqnarray}
  &&S_{(3)}^{2}= \int d^{d}x\sqrt{g}\left\{24K^{2\mu\nu}-16JK^{\mu\nu}
  -4K^{2}g^{\mu\nu}\right\}\partial_{\mu}\varphi_{(3)}
  \partial_{\nu}\varphi_{(3)} , \label{46}\\
  && S_{(3)}^{\Delta_{(3)}}=4\Delta_{(3)}\int
  d^{d}x\sqrt{g}\left\{J^{3}-3K^{2}J +
  2K^{3}\right\}\varphi_{(3)}^{2} .
\end{eqnarray}
Now to cancel the second line in (\ref{cotton}) with the Cotton tensor
we have to turn to the Bach tensor transformation (\ref{bachtr}). It
is easy to see that the following combination
\begin{equation}\label{bpart}
    S_{(3)}^{B}=-\frac{8}{d-4}\int d^{d}x\sqrt{g}\left\{B^{\mu\nu}
    \partial_{\mu}\varphi_{(3)}\partial_{\nu}\varphi_{(3)}
    +\Delta_{(3)}B^{\mu\nu}K_{\mu\nu}\varphi_{(3)}^{2}\right\} ,
\end{equation}
make our action completely locally conformal invariant. It follows that
that the required locally Weyl invariant action for the $k=3$ case is
\begin{equation}\label{final}
    S_{(3)}=\sum^{2}_{i=0}S_{(3)}^{i}+S_{(3)}^{\Delta_{(3)}} +
    S_{(3)}^{B} .
\end{equation}
Now we analyze the linear on $\Delta_{(3)}\varphi_{(3)}^{2}$ part of
(\ref{final}):
\begin{equation}\label{hol}
   4\Delta_{(3)}\int
  d^{d}x\sqrt{g}\left\{J^{3}-3K^{2}J + 2K^{3}-\frac{2}{d-4}
  B^{\mu\nu}K_{\mu\nu}\right\}\varphi_{(3)}^{2} .
\end{equation}
We see again that this part coincides with the so-called
"holographic" anomaly \cite{HS} in 6 dimensions written in
general spacetime dimension $d$ ( see  also \cite{OA} for the role
of the Bach tensor in holography ). The main property of the holographic
anomaly is that it is a special combination of the Euler density
with the other three Weyl invariants \cite{Deser:1996na} which
reduce the topological part of the anomaly to the expression (\ref{hol})
(see \cite{Bastianelli:2000hi} for the correct separation), which is
zero for the Ricci flat metric.

But this is for the anomaly itself in $d=6$. Here we are concerned
with the invariant Lagrangian and presence of the scalar field and the
integral make our considerations easier. To get the invariant action with
the whole third Euler density, we have to perform some more work,
and find that there is another invariant action with the maximum of
four derivatives.
This action can be obtained, using the same Noether
procedure, to render the following initial term
\begin{equation}\label{Cpart}
    S_{W}^{0}=\frac{8}{(d-3)(d-4)}\int d^{d}x\sqrt{g}
    W^{\mu\alpha\nu\beta}\widehat{D}^{(3)}_{\mu\nu}
    \varphi_{(3)}\widehat{D}^{(3)}_{\alpha\beta}\varphi_{(3)}
\end{equation}
invariant. After some lengthy but straightforward calculation we arrive at
the following locally conformal invariant action.
\begin{equation}\label{w2}
    S_{W}=S_{(3)}^{B}-S_{W}^{0}-S_{W}^{1}-S_{W}^{\Delta_{(3)}}, \quad \delta
    S_{W}=0 ,
\end{equation}
where
\begin{eqnarray}
   S_{W}^{1}&=&\int d^{d}x\sqrt{g}\left\{\frac{16W^{\mu\alpha\nu\beta}
  K_{\alpha\beta}}{(d-3)}
  + \frac{3W^{2}g^{\mu\nu}-12W^{2\mu\nu}}{(d-3)(d-4)}
  \right\} \partial_{\mu}\varphi_{(3)}\partial_{\nu}\varphi_{(3)} ,\label{53}\\
  S_{W}^{\Delta_{(3)}}&=&\Delta_{(3)}\int d^{d}x\sqrt{g}\left\{\frac{12W^{\mu\alpha\nu\beta}
  K_{\mu\nu}K_{\alpha\beta}}{(d-3)}
  + \frac{3W^{2}J-12W^{2\mu\nu}K_{\mu\nu}}{(d-3)(d-4)}
  \right\}\varphi_{(3)}^{2} .\quad\label{54}
\end{eqnarray}
To derive this we used the Bianchi identity (\ref{18}) contracted
with the Weyl tensor. This leads to the following relation
\begin{equation}\label{bi3}
    \frac{1}{2}\partial_{\alpha}W^{2}-2\nabla_{\mu}W^{2\mu}_{\alpha}
    =2(d-4)C^{\quad\nu}_{\lambda\rho}W_{\nu\alpha}^{\quad\lambda\rho}\,,
\end{equation}
which generates the terms quadratic in the Weyl tensor in
(\ref{w2})-(\ref{54}). Therefore the existence of the invariant
(\ref{w2}) allows us to replace the Bach tensor
dependent term $S^{B}_{(3)}$ in (\ref{final}) with $W$ dependent terms and obtain
\begin{equation}\label{finalw}
    S_{(3)}^{\mathcal{A}}=\sum^{2}_{i=0}S^{i}_{(3)}
    +S^{0}_{W}+S^{1}_{W}+S^{\Delta_{(3)}}_{(3)}+S^{\Delta_{(3)}}_{W}
    .
\end{equation}
Then we see that all terms proportional to
$\Delta_{3}\varphi_{(3)}^{2}$ are accumulated in the last two terms
of (\ref{finalw})
\begin{equation}\label{anom}
    S^{\Delta_{(3)}}_{(3)}+S^{\Delta_{(3)}}_{W}=\frac{3\Delta_{3}}{(d-5)(d-4)(d-3)}\int
    d^{d}x\sqrt{g}\mathcal{A}\varphi_{(3)}^{2} ,
\end{equation}
where
\begin{eqnarray}
  \mathcal{A}&=&(d-5)
  [W^{2}J-4W^{2\mu\nu}K_{\mu\nu}]
  +4(d-5)(d-4)W^{\mu\alpha\nu\beta}K_{\mu\nu}
  K_{\alpha\beta}\nonumber\\ &&+\frac{4}{3}(d-5)(d-4)(d-3)
  [J^{3}-3K^{2}J + 2K^{3}] .
\end{eqnarray}
We can now insert (\ref{12}) in (\ref{Ek}) for $k=3$ and get
\begin{eqnarray}
   E_{(3)}&=&\frac{16}{3} W^{3}+\frac{32}{3} W^{\tilde{3}}+ 8\mathcal{A} ,\\
   W^{3}&=&W^{\alpha\beta}_{\mu\nu}W^{\mu\nu}_{\lambda\rho}
   W^{\lambda\rho}_{\alpha\beta}\quad\,,
   W^{\tilde{3}}=W_{\alpha\mu\nu\beta}W^{\lambda\mu\nu\rho}
   W^{\alpha\,\,\,\,\,\beta}_{\,\,\,\,\lambda\rho} .
\end{eqnarray}
In the same way as in the $k=2$ case we can add
these two additional invariant actions with the appropriate
coefficients:
\begin{eqnarray}
  && S_{W^{3}}=\frac{1}{2}\int
    d^{d}x\sqrt{g}\frac{4\Delta_{3}(W^{3}+2 W^{\tilde{3}})}{(d-5)(d-4)(d-3)}
    \varphi_{(3)}^{2} ,
\end{eqnarray}
and restore the Euler density containing Lagrangian
\begin{eqnarray}\label{eds}
   S_{(3)}^{E_{(3)}}&=& S_{(3)}^{\mathcal{A}}+ S_{W^{3}}\nonumber\\
   &=&\frac{1}{2}\int d^{d}x\sqrt{g}
  \left\{-\varphi_{(2)}\Box^{3}\varphi_{(2)}+\,\,\dots\,\,
  +\frac{3\Delta_{3}}{4(d-5)(d-4)(d-3)}E_{(3)}\varphi_{(3)}^{2}\right\} ,\quad\quad
\end{eqnarray}
where we put $\dots$ instead of the other terms with derivatives, or
terms proportional to $\Delta^{2}_{(3)}$ and $\Delta^{3}_{(3)}$.
These terms can be readily read off (\ref{ks0}), (\ref{44}),
(\ref{Cpart}) and (\ref{53}).

We have proved our assertion concerning the connection between the hierarchy
of conformally coupled scalars with the dimensions $\Delta_{k}$  and
Euler densities $E_{(k)}$ for the $k=1,2,3$, and have constructed the
conformal coupling of the third scalar with gravity in dimensions
$d\geq 6$. This action in spacetime dimension $d=6$ or
equivalently for $\Delta_{(3)}=0$ degenerates to a conformal invariant
operator for dimension $0$ scalars obtained in \cite{KMM,AKMM} from
cohomological considerations of the effective action.

\section*{Conclusion}

In conclusion, we formulate our assertion for general $k$ case. Comparing
(\ref{act11}), (\ref{se}) and (\ref{eds}) we expect the
following terms in the action of conformally coupled  scalar with
the scaling dimensions $\Delta_{k}=k-d/2$
\begin{eqnarray}\label{efinal}
   &&S_{(k)}^{E_{(k)}}=\frac{(-1)^{k}}{2}\int d^{d}x\sqrt{g}
  \left\{\varphi_{(k)}\Box^{k}\varphi_{(k)}+\,\,\dots\,\,
  -\frac{k!(d-2k)!\Delta_{k}}{2^{k}(d-k)!}E_{(k)}\varphi_{(k)}^{2}\right\}\,.\quad\quad
\end{eqnarray}

\subsection*{Acknowledgements}
\quad We are very grateful to Eugen Radu for numerous thoughtful
discussions on this subject. Our thanks are due also  Werner Nahm,
Denjoe O'Connor and Ivo Sachs for helpful comments.\quad
This work is carried out in the framework of Enterprise--Ireland
Basic Science Research Project SC/2003/390 of Enterprise-Ireland.
\quad The work of RM is partially supported by INTAS grant
\#03-51-6346.

%
%
%

\end{document}